\documentstyle[graphics,aps,twocolumn,prb]{revtex}
\begin{document}
\draft

\title{Energy dissipation and scattering angle distribution analysis of the
  classical trajectory calculations of methane scattering from a Ni(111) surface}

\author{Robin Milot}
\address{Schuit Institute of Catalysis, ST/SKA, Eindhoven University of
  Technology,\\ P.O. Box 513, NL-5600 MB Eindhoven, The Netherlands.}

\author{A.W. Kleyn}
\address{Leiden Institute of Chemistry, 
   Department of Surfaces and Catalysis, Leiden University,\\ P.O. Box
   9502, NL-2300 RA Leiden, The Netherlands.}

\author{A.P.J. Jansen}
\address{Schuit Institute of Catalysis, ST/SKA, Eindhoven University of
  Technology.}

\date{\today}
\maketitle

\begin{abstract}
  We present classical trajectory calculations of the rotational
  vibrational scattering of a non-rigid methane molecule from a Ni(111)
  surface. Energy dissipation and scattering angles have been studied as
  a function of the translational kinetic energy, the incidence angle,
  the (rotational) nozzle temperature, and the surface temperature.
  Scattering angles are somewhat towards the surface for the incidence
  angles of 30${}^{\circ}$, 45${}^{\circ}$, and 60${}^{\circ}$ at a
  translational energy of 96 kJ/mol. Energy loss is primarily from the
  normal component of the translational energy. It is transfered for
  somewhat more than half to the surface and the rest is transfered
  mostly to rotational motion.  The spread in the change of
  translational energy has a basis in the spread of the transfer to
  rotational energy, and can be enhanced by raising of the surface
  temperature through the transfer process to the surface motion.
\end{abstract}

\section{Introduction}
\label{dyn:intro}

The dissociative adsorption of methane on transition metals is an
important reaction in catalysis; it is the rate limiting step in steam
reforming to produce syngas, and it is prototypical for catalytic C--H
activation. Therefore the dissociation is of high interest for many
surface scientists. (See for a recent review Ref.~\onlinecite{lar00}.)
Molecular beam experiments in which the dissociation probability was
measured as a function of translational energy have observed that the
dissociation probability is enhanced by the normal incidence component of
the incidence translational
energy.\cite{ret85,ret86,lee87,hol96,xxx13,hol95,lun89,walker99,walker00,see97,see97b}
This suggests that the reaction occurs primarily through a direct
dissociation mechanism at least for high translational kinetic energies.
Some experiments have also observed that vibrationally hot CH${}_4$
dissociates more readily than cold CH${}_4$, with the energy in the
internal vibrations being about as effective as the translational energy
in inducing
dissociation.\cite{ret85,ret86,lee87,lun89,hol95,lar99,walker99,walker00}
A molecular beam experiment with laser excitation of the $\nu_3$ mode
did succeed in measuring a strong enhancement of the dissociation on a
Ni(100) surface. However, this enhancement was still much too low to
account for the vibrational activation observed in previous studies and
indicated that other vibrationally excited modes contribute
significantly to the reactivity of thermal samples.\cite{juur99}

It is very interesting to simulate the dynamics of the dissociation,
because of the direct dissociation mechanism, and the role of the
internal vibrations. Wave packet simulations of the methane dissociation
reaction on transition metals have treated the methane molecule always
as a diatomic up to now.\cite{har91,lun91,lun92,lun95,jan95,car98} Apart
from one C--H bond (a pseudo $\nu_3$ stretch mode) and the molecule
surface distance, either (multiple) rotations or some lattice motion
were included. None of these studies have looked at the role of the
other internal vibrations, so there is no model that describes which
vibrationally excited mode might be responsible for the experimental
observed vibrational activation.

In previous papers we have reported on wave packet simulations to
determine which and to what extent internal vibrations are important for
the dissociation in the vibrational ground state of
CH${}_4$,\cite{mil98} and CD${}_4$.\cite{mil00a} We were not able yet to
simulate the dissociation including all internal vibrations. Instead we
simulated the scattering of methane in fixed orientations, for which all
internal vibrations can be included, and used the results to deduce
consequences for the dissociation. These simulations indicate that to
dissociate methane the interaction of the molecule with the surface
should lead to an elongated equilibrium C--H bond length close to the
surface, and that the scattering was almost elastic. Later on we
reported on wave packet simulations of the role of vibrational
excitations for the scattering of CH${}_4$ and CD${}_4$.\cite{mil00b} We
predicted that initial vibrational excitations of the asymmetrical
stretch ($\nu_3$) but especially the symmetrical stretch ($\nu_1$) modes
will give the highest enhancement of the dissociation probability of
methane.  Although we have performed these wave packet simulations in
ten dimensions, we still had to neglect two translational and three
rotational coordinates of the methane molecule and we did not account
for surface motion and corrugation. It is nowadays still hard to include
all these features into a wave packet simulation, therefore we decided
to study these with classical trajectory simulations.

In this article we will present full classical trajectory simulations of
methane from a Ni(111) surface. We have especially interest in the
effect of the molecular rotations and surface motion, which we study as
a function of the nozzle and surface temperature. The methane molecule
is flexible and able to vibrate. We do not include vibrational kinetic
energy at the beginning of the simulation, because a study of
vibrational excitation due to the nozzle temperature needs a special
semi-classical treatment. Besides its relevance for the dissociation
reaction of methane on transition metals, our scattering simulation can
also be of interest as a reference model for the interpretation of
methane scattering itself, which have been studied with molecular beams
on Ag(111),\cite{asada81,asada82}
Pt(111),\cite{yagyu99,yagyu99b,yagyu00,hiraoka00} and Cu(111)
surfaces.\cite{Andersson00} It was observed that the scattering angles
are in some cases in disagreement with the outcome of the classical Hard
Cube Model (HCM) described in
Ref.~\onlinecite{logan66}.\cite{yagyu99,yagyu99b} We will show in this
article that the assumptions of this HCM model are too crude for
describing the processes obtained from our simulation. The
time-of-flight experiments show that there is almost no vibrational
excitation during the scattering,\cite{yagyu00,hiraoka00} which is in
agreement with our current classical simulations and our previous wave
packet simulations.\cite{mil98,mil00a}
 
The rest of this article is organized as follows. We start with a
description of our model potential, and an explanation of the 
simulation conditions. The results and discussion are presented next. We
start with the scattering angles, and relate them to the energy
dissipation processes. Next we will compare our simulation with other
experiments and theoretical models. We end with a summary and some
general conclusions.

\section{Computational details}
\label{dyn:comp}

We have used classical molecular dynamics for simulating the scattering
of methane from a Ni(111) surface. The methane molecule was modelled as
a flexible molecule. The forces on the carbon, hydrogen, and Ni atoms
are given by the gradient of the model potential energy surface
described below.  The first-order ordinary differential equations for
the Newtonian equations of motion of the Cartesian coordinates were
solved with use of a variable-order, variable-step Adams
method.\cite{nag17} We have simulated at translational energies of 24,
48, 72, and 96 kJ/mol at normal incidence, and at 96 kJ/mol for incidence
angles of 30${}^{\circ}$, 45${}^{\circ}$, and 60${}^{\circ}$ with the
surface normal. The surface temperature and (rotational) nozzle temperature for
a certain simulation were taken independently between 200 and 800 K.

\subsection{Potential energy surface}
\label{dyn:pes}

The model potential energy surface used for the classical dynamics is
derived from one of our model potentials with elongated C--H bond
lengths towards the surface, previously used for wave packet simulation
of the vibrational scattering of fixed oriented methane on a flat
surface.\cite{mil98,mil00a} In this original potential there is one part
responsible for the repulsive interaction between the surface and the
hydrogens, and another part for the intramolecular interaction between
carbon and hydrogens.

We have rewritten the repulsive part in pair potential terms between top
layer surface Ni atoms and hydrogens in such a way that the surface
integral over all these Ni atoms give the same overall exponential
fall-off as the original repulsive PES term for a methane molecule far
away from the surface in an orientation with three bonds pointing
towards the surface.  The repulsive pair interaction term $V_{\rm rep}$
between hydrogen $i$ and Ni atom $j$ at the surface is then given by
\begin{equation}
  \label{Vrep}
  V_{\rm rep}=\frac{A\ e^{-\alpha Z_{ij}}} {Z_{ij}},
\end{equation}
where $Z_{ij}$ is the distance between hydrogen atom $i$ and Ni atom $j$.

The intramolecular potential part is split up in bond, bond angle, and
cross potential energy terms. The single C--H bond energy is given
by a Morse function with bond lengthening towards the surface
\begin{equation}
  \label{Vbond}
  V_{\rm bond}=  D_e \ \Big[1-e^{ -\gamma (R_i-R_{eq})}\Big]^2 ,
\end{equation}
where $D_e$ is the dissociation energy of methane in the gas phase, and
$R_i$ is the length of the C--H bond $i$. Dissociation is not possible
at the surface with this potential term, but the entrance channel for
dissociation is mimicked by an elongation of the equilibrium bond length
$R_{eq}$ when the distance between the hydrogen atom and the Ni atoms in
the top layer of the surface become shorter. This is achieved by
\begin{equation}
  \label{Req}
  R_{eq}= R_0 + S~\sum\limits_{j} \frac{e^{-\alpha Z_{ij}}} {Z_{ij}} ,
\end{equation}
where $R_0$ is the equilibrium C--H bond length in the gas phase. The
bond elongation factor $S$ was chosen in such a way that the elongation
is 0.054 nm at the classical turning point of 93.2 kJ/mol incidence
translational energy for a rigid methane molecule, when the molecule
approach a surface Ni atom atop with one bond pointing towards the
surface.  The single angle energy is given by the harmonic expression
\begin{equation}
  \label{Vangle}
  V_{\rm angle} = k_{\theta}\ (\theta_{ij} - \theta_{0})^2 ,
\end{equation}
where $\theta_{ij}$ is the angle between C--H bond $i$ and $j$, and
$\theta_{0}$ the equilibrium bond angle.
Furthermore, there are some cross-term potentials between bonds and
angles. The interaction between two bonds are given by
\begin{equation}
  \label{Vbb}
  V_{\rm bb} = k_{RR}\ (R_i-R_0)(R_j-R_0) .
\end{equation}
The interaction between a bond angle and the bond angle on the other side is
given by
\begin{equation}
  \label{Vaa}
  V_{\rm aa} = k_{\theta\theta}\ 
             (\theta_{ij}-\theta_{0})(\theta_{kl}- \theta_{0}).
\end{equation}
The interaction between a bond angle and one of its bonds is given by
\begin{equation}
  \label{Vab}
  V_{\rm ab} = k_{\theta R}\ (\theta_{ij}-\theta_{0})(R_i- R_{0}).
\end{equation}
The parameters of the intramolecular potential energy terms
were calculated by fitting the second derivatives of these terms on the
experimental vibrational frequencies of CH${}_4$ and
CD${}_4$ in the gas phase.\cite{gray79,lee95}

The Ni-Ni interaction between nearest-neighbours is given by the
harmonic form
\begin{eqnarray}
  \label{Vnini}
  V_{\rm Ni-Ni} & = & \frac{1}{2} \lambda_{ij} [(\mathbf{u}_i -
                  \mathbf{u}_j) \cdot \mathbf{\hat r}_{ij} ]
                  \nonumber\\ 
                  & & + \frac{1}{2} \mu_{ij}  \Big\{(\mathbf{u}_i -
                  \mathbf{u}_j)^2 - [(\mathbf{u}_i -
                  \mathbf{u}_j) \cdot \mathbf{\hat r}_{ij} ]^2 \Big\}.
\end{eqnarray}
The $\mathbf{u}$'s are the displacements from the equilibrium positions,
and $\mathbf{\hat r}$ is a unit vector connecting the equilibrium
positions. The Ni atoms were placed at bulk positions with a
nearest-neighbour distance of 0.2489 nm. The parameters $\lambda_{ij}$
and $\mu_{ij}$ were fitted on the elastic constants \cite{Landolt} and
cell parameters \cite{Ashcroft} of the bulk. The values of all parameters
are given in Table \ref{tab:traj_par}.

\subsection{Simulation model}

The surface is modelled by a slab consisting of four layers of eight
times eight Ni atoms. Periodic boundary conditions have been used in the
lateral direction for the Ni-Ni interactions. The methane molecule has
interactions with the sixty-four Ni atoms in the top layer of the slab.
The surface temperature is set according to the following procedure. The
Ni atoms are placed in equilibrium positions and are given random
velocities out of a Maxwell-Boltzmann distribution with twice the
surface temperature. The velocities are corrected such that the total
momentum of all surface atoms is zero in all directions, which fixes the
surface in space. Next the surface is allowed to relax for 350 fs.
We do the following ten times iteratively. If at the end of previous
relaxation the total kinetic energy is above or below the given surface
temperature, then all velocities are scaled down or up with a factor of
$\sqrt 1.1$ respectively. Afterwards a new relaxation simulation is
performed. The end of each relaxation run is used as the begin condition
of the surface for the actual scattering simulation.

The initial perpendicular carbon position was chosen 180 nm above the
equilibrium $z$-position of the top layer atoms and was given randomly
parallel ($x$, $y$) positions within the central surface unit cell of
the simulation slab for the normal incidence simulations. The methane
was placed in a random orientation with the bonds and angles of the
methane in the minimum of the gas phase potential. The initial
rotational angular momentum was generated randomly from a
Maxwell-Boltzmann distribution for the given nozzle temperature for all
three rotation axis separately. No vibrational kinetic energy was given
initially. Initial translational velocity was given to all methane
atoms according to the translational energy. The simulations under an
angle were given parallel momentum in the [110] direction. The parallel
positions have been translated according to the parallel velocities in
such a way that the first collision occurs one unit cell before the
central unit cell of the simulation box. We tested other directions, but
did not see any differences for the scattering.

Each scattering simulation consisted of 2500 trajectories with a
simulation time of 1500 fs each. We calculated the (change of)
translational, total kinetic, rotational and vibrational kinetic,
intramolecular potential, and total energy of the methane molecule; and
the scattering angles at the end of each trajectory. We calculated for
them the averages and standard deviations, which gives the spread for
the set of trajectories, and correlations coefficients from which we can
abstract information about the energy transfer processes.

\section{Results and discussion}
\label{dyn:results}

We will now present and discuss the results of our simulations. We begin
with the scattering angle distribution. Next we will explain this in
terms of the energy dissipation processes. Finally we will compare our
simulation with previous theoretical and experimental scattering
studies, and discuss the possible effects on the dissociation of methane
on transition metal surfaces.

\begin{table}
  \caption{Parameters of the potential energy surface.}
  \label{tab:traj_par}
  \begin{tabular}{l l r l}
    Ni--H & $A$ & 971.3 & kJ nm mol${}^{-1}$ \\
     & $\alpha$ & 20.27   & nm${}^{-1}$ \\
     & $S$      & 0.563     & nm${}^2$  \\
     \\
    CH${}_4$    & $\gamma$ & 17.41 & nm${}^{-1}$\\
     & $D_e$    & 480.0 & kJ mol${}^{-1}$\\
     & $R_0$    & 0.115 & nm \\
     & $k_{\theta}$ & 178.6 & kJ mol${}^{-1}$ rad${}^{-2}$ \\
     & $\theta_0$   & 1.911 & rad \\
     & $k_{RR}$     & 4380 & kJ mol${}^{-1}$ nm${}^{-2}$ \\
     & $k_{\theta\theta}$ & 11.45 & kJ mol${}^{-1}$ rad${}^{-2}$\\
     & $k_{\theta R}$ & -472.7 & kJ mol${}^{-1}$ rad${}^{-1}$ nm${}^{-1}$\\
     \\
    Ni--Ni & $\lambda_{nn}$ & 28328 & kJ mol${}^{-1}$ nm${}^{-2}$ \\
     & $\mu_{nn}$ & -820 & kJ mol${}^{-1}$ nm${}^{-2}$ \\
  \end{tabular}
\end{table}

\subsection{Scattering angles}

Figure \ref{fig:sctangdst} shows the scattering angle distribution for
different incidence angles with a initial total translational energy of
96 kJ/mol at nozzle and surface temperatures of both 200 and 800 K. The
scatter angle is calculated from the ratio between the normal and the
total parallel momentum of the whole methane molecule. We observe that
most of the trajectories scatter some degrees towards the surface from
the specular. This means that there is relatively more parallel momentum
than normal momentum at the end of the simulation compared with the
initial ratio. This ratio change is almost completely caused by a
decrease of normal momentum.

The higher nozzle and surface temperatures have almost no influence on
the peak position of the distribution, but give a broader distribution.
The standard deviation in the scattering angle distribution goes up from
$2.7^{\circ}$, $2.4^{\circ}$, and $2.2^{\circ}$ at 200K to
$4.4^{\circ}$, $3.8^{\circ}$, and $3.4^{\circ}$ at 800K for incidence
angles of $30^{\circ}$, $45^{\circ}$, and $60^{\circ}$ respectively.
This means that the angular width is very narrow, because the full width
at half maximum (FWHM) are usually larger than
$20^{\circ}$.\cite{wiskerke95} (The FWHM is approximately somewhat more
than twice the standard deviation.)  The broadening is caused almost
completely by raising the surface temperature, and has again
primarily an effect on the spread of the normal momentum of the
molecule. This indicates that the scattering of methane from Ni(111) is
dominated by a thermal roughening process.

We do not observe an average out-of-plane diffraction for the non normal
incidence simulations, but we do see some small out-of-plane
broadening. The standard deviations in the out-of-plane angle were
0.9${}^{\circ}$, 1.8${}^{\circ}$, 3.4${}^{\circ}$ at a surface
temperature of 200K, and 1.7${}^{\circ}$, 3.3${}^{\circ}$, and
6.0${}^{\circ}$ at 800K for incidence angles of 30${}^{\circ}$,
45${}^{\circ}$, and 60${}^{\circ}$ with the surface normal. Raising the
(rotational) nozzle temperature has hardly any effect on the
out-of-plane broadening.

\subsection{Energy dissipation processes}

\subsubsection{Translational energy}

\begin{figure}
\includegraphics{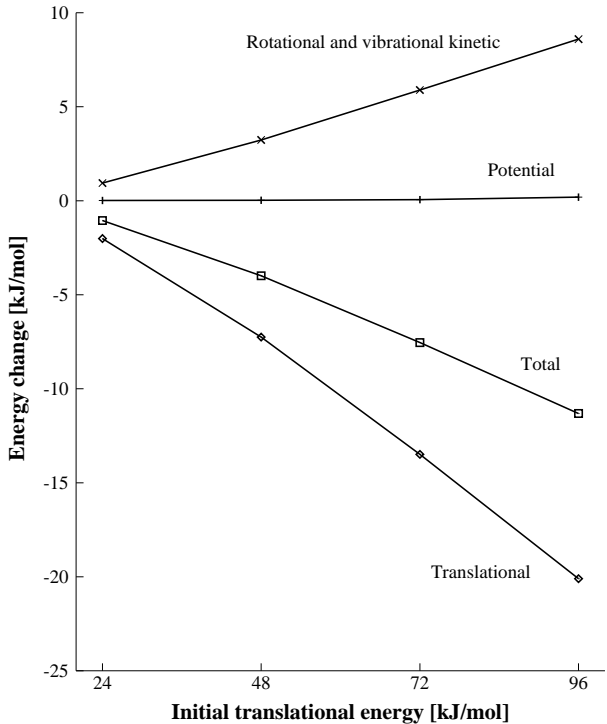}%
\caption{The average energy change  (kJ/mol) of the methane translational
  energy, the  methane total energy, the methane potential energy,
  and the methane rotational and vibrational kinetic energy as a function
  translational kinetic energy (kJ/mol) at normal incidence. The nozzle
  and surface temperature were 400K. }
\label{fig:dtotav}
\end{figure}

Figure \ref{fig:dtotav} shows the average energy change of some energy
components of the methane molecule between the end and the begin of the
trajectories as a function of the initial total translational energy.
The incoming angle for all is 0${}^{\circ}$ (normal incidence), and both
the nozzle and surface are initially 400K. If we plot the normal incidence
translational energy component of the simulation at 96 kJ/mol for the
different incidence angles, then we see a similar relation. This means
that there is normal translational energy scaling for the scattering
process in general, except for some small differences discussed later
on.

\begin{figure}
\includegraphics{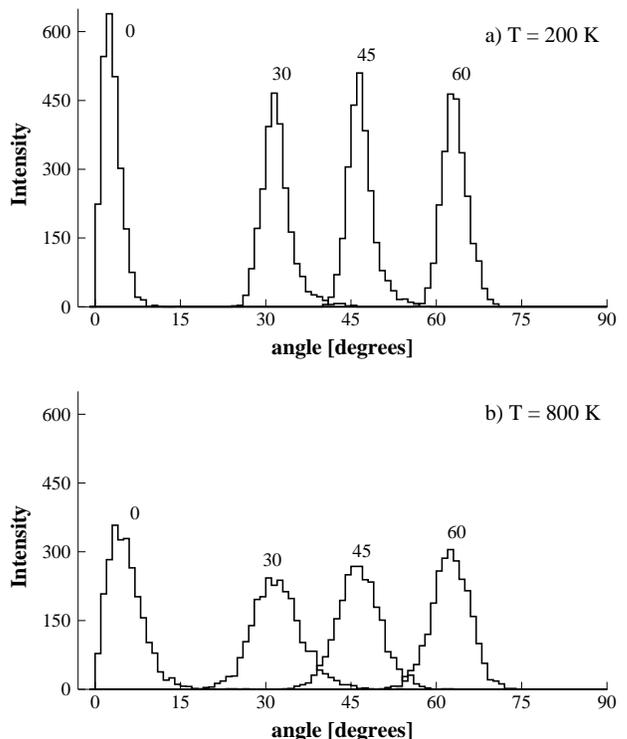}%
\caption{The distribution of the scattering angle for a total initial
  translational energy of 96 kJ/mol with incidence angles of
  0${}^{\circ}$, 30${}^{\circ}$, 45${}^{\circ}$, and 60${}^{\circ}$ with
  the surface normal. Both the nozzle and surface temperature are: a) 200K,
  and b) 800K.}
\label{fig:sctangdst}
\end{figure}

Most of the initial energy of methane is available as translational
energy, so it cannot be surprising that we see here the highest energy
loss. The translational energy loss takes a higher percentage of the
initial translational energy at higher initial translational energies.
Since almost all of the momentum loss is in the normal direction, we
also see that the loss of translational energy can be found back in the
normal component of the translational energy for the non-normal incidence
simulations.

The average change of the total energy of the methane molecule is less
negative than the average change in translational energy, which 
means that there is a net transfer of the initial methane energy towards
the surface during the scattering. This is somewhat more than half of
the loss of translational energy. The percentage of transfered energy to
the surface related to the normal incidence translational energy is also
enhanced at higher incidence energies.  There is somewhat more
translational energy loss, and energy transfer towards the surface for
the larger scattering angles, than occurs at the comparable normal
translational energy at normal incidence.  This is caused probably by
interactions with more surface atoms, when the molecule scatters under
an larger angle with the surface normal.

In Fig.~\ref{fig:dtotav} we also plotted the average change of methane
potential energy and the change of rotational and vibrational kinetic
energy of methane.  We observe that there is extremely little energy
transfer towards the potential energy, and a lot of energy transfer
towards rotational and vibrational kinetic energy. Vibrational motion
gives an increase of both potential and kinetic energy. Rotational
motion gives only an increase in kinetic energy. So this means that
there is almost no vibrational inelastic scattering, and very much
rotational inelastic scattering.

\begin{figure}
\includegraphics{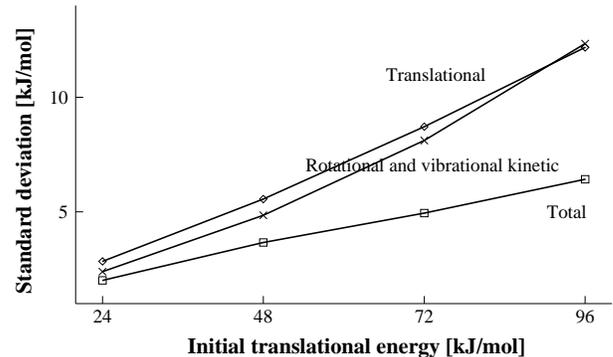}%
\caption{The standard deviation in the energy change (kJ/mol) of the
  methane translational energy, the methane total energy, and the methane
  rotational and vibrational kinetic energy as a function of the initial
  translational energy (kJ/mol) at normal incidence. The surface and
  nozzle temperature are both 200K. }
\label{fig:st_dev_trans}
\end{figure}

Figure \ref{fig:st_dev_trans} shows the standard deviations in the
energy change of some energy components of methane versus the initial
translational energy at normal incidence for a nozzle and surface
temperature of 200K. (The temperature effects will be discussed below.)
The standard deviations in the energy changes are quite large compared
to the average values. The standard deviations in the change of the
methane translational energy and in the change of methane rotational and
vibrational kinetic energy increase more than the standard deviation in the
change of methane total energy, when the initial translational energy is
increased.  We find again an identical relation if we plot the standard
deviations versus the initial normal energy component of the scattering
at different incidence angles. The standard deviations are much smaller
in the parallel than in the normal component of the translational
energy, so again only the normal component of the translational energy
is important.  Although the standard deviations in the translational
energy is smaller at larger incidence angles than at smaller incidence
angles, we see in Fig.~\ref{fig:sctangdst} that the spread in the angle
distribution is almost the same. This is caused by the fact that at
large angles deviations in the normal direction has more effect on the
deviation in the angle than at smaller angles with the normal.

\subsubsection{Surface temperature}

An increase of surface temperature gives a small reduction of average
translational energy loss (around 5 $\%$ from 200K to 800K at 96 kJ/mol
normal incidence). This is the reason why we do not observe a large shift
of the peak position of the scattering angle distribution. However, an
increase of surface temperature does have a larger effect on the average
energy transfer to the surface, but this is in part compensated through
a decrease of energy transfer to rotational energy.

\begin{figure}
\includegraphics{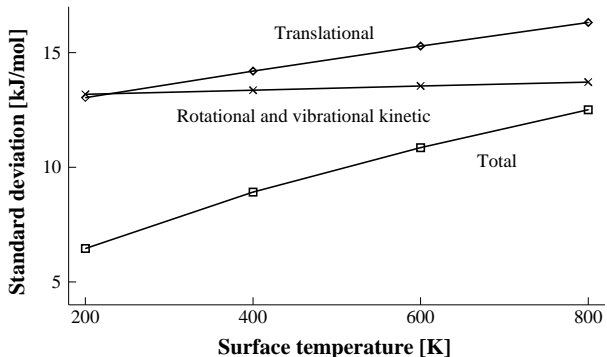}%
\caption{The standard deviation in the energy change (kJ/mol) of the
  methane translational energy, the methane total energy, and the methane
  rotational and vibrational kinetic energy as a function of the surface
  temperature (K). The nozzle temperature is 400K, and the translational
  energy is 96 kJ/mol at normal incidence. }
\label{fig:st_dev_surfT}
\end{figure}

Figure \ref{fig:st_dev_surfT} shows the standard deviations in the
energy change of the translational energy, the methane total energy, and
the methane rotational and vibrational kinetic energy as a function of
the surface temperature. We observe that the standard deviation in the
change of rotational and vibrational kinetic energy hardly changes at
increasing surface temperature. At a low surface temperature it is much
higher than the standard deviation in the change of the methane total
energy. So the baseline broadening of translational energy is caused by
the transfer of translational to rotational motion.  The standard
deviation in the change of the methane total energy increases much at
higher surface temperature. This results also in an increase of the
standard deviation in the change of translational energy, which means
that the surface temperature influences the energy transfer process
between translational and surface motion.  The spread in the change of
translational energy is related to the spread in the scattering angle
distributions. It is now clear that the observed broadening of the
scattering angle distribution with increasing surface temperature is
really caused by a thermal roughening process.

\subsubsection{Nozzle temperature}

\begin{figure}
\includegraphics{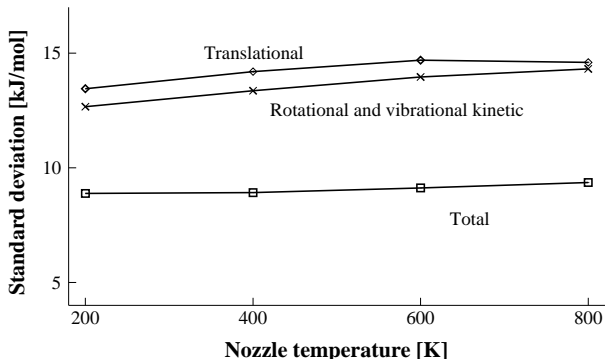}%
\caption{The standard deviation in the energy change (kJ/mol) of the
  methane translational energy, the methane total energy, and the methane
  rotational and vibrational kinetic energy as a function of the nozzle
  temperature (K). The surface temperature is 400K, and the translational
  energy is 96 kJ/mol at normal incidence. }
\label{fig:st_dev_nozzleT}
\end{figure}

Figure \ref{fig:st_dev_nozzleT} shows the dependency of the standard
deviations for the different energy changes on the nozzle temperature.
From this figure it is clear that the nozzle temperature has relative
little influence on the standard deviations in the different energy
changes. Therefore we observe almost no peak broadening in the
scattering angle distribution due to the nozzle temperature.
The nozzle temperature has also no influence on the average change of
rotational and vibrational kinetic energy, which means that this part of the
energy transfer process is driven primarily by normal incidence
translational energy.

We have to keep in mind that we only studied the rotational heating by
the nozzle temperature, and that we did not take vibrational excitation
by nozzle heating into account. From our wave packet simulations we know
that vibrational excitations can contribute to a strong enhancement of
vibrational inelastic scattering.\cite{mil00b} So the actual effect of
raising the nozzle temperature can be different than sketched here.  


\subsection{Comparison with other studies}

\subsubsection{Scattering angles and the Hard Cube Model}

The angular dependnece of scattered intensity for a fixed total
scattering angle has only been measured at
Pt(111).\cite{yagyu99,yagyu99b} The measurement has been compared with
the predictions of the Hard Cube Model (HCM) as described in
Ref.~\onlinecite{logan66}. There seems to be more or less agreement for
low translational energies under an angle around 45${}^{\circ}$ with the
surface, but is anomalous at a translational energy of 55 kJ/mol. The
anomalous behaviour has been explained by altering the inelastic
collision dynamics through intermediate methyl fragments.

Although our simulations are for Ni(111) instead of Pt(111) and we
calculate real angular distributions, we will show now that the HCM is
insufficient for describing the processes involved with the scattering
of methane in our simulation. The HCM neglects the energy transfer to
rotational excitations, and overestimates the energy transfer to the
surface. This is not surprising, because the HCM is constructed as a
simple classical model for the scattering of gas atoms from a solid
surface. The basic assumptions are that (1) the interaction of the gas
atom with a surface atom is represented by an impulsive force of
repulsion, (2) the gas-surface intermolecular potential is uniform in
the plane of the surface, (3) the surface is represented by a set of
independent particles confined by square well potentials, (4) the
surface particles have a Maxwellian velocity distribution.\cite{logan66}
Assumption 1 excludes inelastic rotational scattering, because the gas
particle is an atom without moment of inertia.  So the HCM misses a
large part of inelastic scattering. However, it still predicts
scattering angles much more below the incidence angles than we found
from our simulation. For example: The HCM predicts an average scattering
angle with the surface normal of 64${}^{\circ}$ from Ni(111), at an
incidence angle of 45${}^{\circ}$ at a surface temperature four times
lower than the gas temperature. This is much more than for Pt(111),
because the mass ratio between the gas particle and the surface atom is
higher for Ni(111).  There are several explanations for this error.
First, the assumption 3 is unreasonable for atomic surfaces with low
atom weight, because the surface atoms are strongly bound to each other.
This means that effectively the surface has a higher mass than
assumed.\cite{grimmelmann80} Second, there is no one-on-one interaction
between surface atom and methane molecule, but multiple hydrogen atoms
interacting with different Ni atoms.  Third, the methane molecule is not
rigid in contrast to assumption 1.  We have followed the energy
distribution during the simulation for some trajectories and find that
the methane molecule adsorbs initial rotational and translational energy
as vibrational energy in its bonds and bond angles when close the
surface, which is returned after the methane moves away from it.

It would be nice to test our model with molecular beam experiment of the
scattering angles on surfaces with relatively low atom weight, which
also try to look at rotational inelastic scattering.

\subsubsection{Wave packet simulations}

Let us now compare the full classical dynamics with our fixed oriented
wave packet simulations,\cite{mil98,mil00a,mil00b} because this was
initial the reason to perform the classical dynamics simulations.  Again
we observe very little vibrational inelastic scattering. This is in
agreement with the observations in the time-of-flight experiments on
Pt(111).\cite{yagyu00,hiraoka00}

Since we used our wave packet simulations to deduce consequences for the
dissociation of methane, we have to wonder whether the observed
inelastic scattering in our classical simulations changes the 
picture of the dissociation in our previous publications.  Therefore we
have to look at what happens at the surface. We did so by following some
trajectories in time. 

We find approximately the same energy rearrangements for the classical
simulations as discussed for the wave packet simulations for the
vibrational groundstate in Refs.~\onlinecite{mil00a} and
\onlinecite{mil00b}.  Again most of the normal translational energy is
transfered to the potential energy terms of the surface repulsion [see
Eq.~\ref{Vrep}].  This repulsive potential energy was only given back to
translational energy in the wave packet simulations, because the
orientations and surface were fixed. For the classical trajectory
simulations presented in this article, the repulsive potential energy is
transfered to translational, rotational, and surface energy through the
inherent force of the repulsive energy terms. We observe almost no
energy transfers to translational energy parallel to the surface, so
exclusion of these translational coordinates in the wave packet
simulations do not effect our deduction on the dissociation. The energy
transfers to the rotational and surface energy during the collision make
it harder for the molecule to approach the surface. This will have a
quantitative effect on the effective bond lengthening near the surface,
but not a qualitative.

The remaining problem deals with the effect of rotational motion on the
dissociation probability and steering. Our first intension was to look
for the favourable orientation at the surface, but from following some
trajectories it is clear that steering does not seem to occur. There
is always some rotational motion, and the molecule leaves the surface
often with another hydrogen pointing towards to surface than when it
approaches the surface. This indicates that multiple bonds have a chance
to dissociate during one collision. However, it will be very speculative
to draw more conclusion on the dissociation of methane based on the
scattering in these classical trajectory simulations. Classical trajectory
simulation with an extension of our potentials with an exit channel for
dissociation can possibly learn us more.

\section{Conclusions}
\label{dyn:concl}

We have performed classical dynamics simulations of the rotational
vibrational scattering of non-rigid methane from a corrugated Ni(111)
surface. Energy dissipation and scattering angles have been studied as a
function of the translational kinetic energy, the incidence angle, the
(rotational) nozzle temperature, and the surface temperature.

We find the peak of the scattering angle distribution somewhat below the
incidence angle of 30${}^{\circ}$, 45${}^{\circ}$, and 60${}^{\circ}$ at
a translational energy of 96 kJ/mol. This is caused by an average energy
loss in the normal component of the translational energy. An increase of
initial normal translational energy gives an enhancement of inelastic
scattering. The energy loss is transfered for somewhat more than half to
the surface and the rest mostly to rotational motion.  The vibrational
scattering is almost completely elastic.

The broadening of the scattering angle distribution is mainly caused by
the energy transfer process of translational energy to rotational
energy. Heating of the nozzle temperature gives no peak broadening.
Heating of the surface temperature gives an extra peak broadening
through thermal roughening of the surface.

The Hard Cube Model seems to be insufficient for describing the
scattering angles of methane from Ni(111), if we compare its assumptions
with the processes found in our simulations.

\section*{acknowledgments}

This research has been financially supported by the Council for Chemical
Sciences of the Netherlands Organization for Scientific Research
(CW-NWO), and has been performed under the auspices of the Netherlands
Institute for Catalysis Research (NIOK).


\end{document}